\documentclass[12pt]{article}

\renewcommand{\title}[1]{%
	\begin{center} \Large \bf #1 \end{center}%
	}
\renewcommand{\author}[3]{%
	\begin{center} #1@#2 \\ %
	  {\small E-mail: \texttt{#3}}%
	\end{center}%
	\addvspace{\baselineskip}%
	}

\begin{document}
\baselineskip 5mm
\title{
Towards Gravity-Gauge-Higgs Unification\footnote{Talk presented by N.M. 
at YITP workshop on ``Quantum Field Theory 2004'', 
Yukawa Institute for Theoretical Physics at Kyoto University, Japan, 
July 13-16, 2004.}
}

\vspace{1cm}

{\large
\centerline{ K. Hasegawa$^{(a)}$,
C. S. Lim$^{(b)}$
and Nobuhito Maru$^{(c)}$}}
\vspace{0.5cm}
\centerline{$^{(a)}${\it Graduate School of Science and Technology, 
Kobe University, Kobe 657-8501, Japan}}
\centerline{$^{(b)}${\it Department of Physics, Kobe University, 
Kobe 657-8501, Japan}}
\centerline{$^{(c)}${\it Theoretical Physics Laboratory, RIKEN, 
Saitama 351-0198, Japan}}

\vspace{0.5cm}

\begin{abstract}
We discuss a possibility to solve the gauge hierarchy problem 
in the framework of Gravity-Gauge-Higgs Unification scenario. 
We have calculated 1-loop correction to the mass of the scalar field, 
which is originated from 55-component of the metric, 
in five dimensional gravity theory with the bulk scalar field 
compactified on $S^1$. 
It is shown that the quadratic divergences are canceled and 
the finite mass is generated by explicit diagrammatic calculations and 
the effective potential calculations. 
\end{abstract}

\vspace{0.5cm}

One of the approaches to solve the gauge hierarchy problem is 
a Gauge-Higgs unification scenario. 
In this scenario, 
Higgs field is identified as extra dimensional components of the gauge field 
in higher dimensional gauge theory. 
In particular, it has been known that 1-loop correction to the Higgs mass 
becomes finite in five dimensional QED compactified on $S^1$ [1].

In this talk, 
motivated by the Gauge-Higgs unification scenario, 
we have discussed a possibility to solve the gauge hierarchy problem 
in the framework of the Gravity-Gauge-Higgs unification scenario 
in which Higgs is identified as extra dimensional components of 
the metric tensor field [2]. 
As a prototypical model, 
we take a five dimensional gravity theory coupled with a bulk scalar field, 
compactified on $S^1$. 
1-loop correction to Higgs mass is explicitly calculated 
in a diagrammatical way. 
We clarified that quadratic divergences are canceled 
only when all KK modes are summed up in the internal loop 
in order to maintain the general coordinate transformation invariance. 
Furthermore, we have obtained a finite mass which is generated 
by the non-local effects. 
This result corresponds to that the finite mass is obtained 
in five dimensional QED compactified on $S^1$ 
only when all KK modes are summed up in the internal loop in order to 
maintain the local gauge symmetry.  
What is nontrivial in this calculation is that the finite mass cannot 
be obtained unless the vacuum bubble diagram and the tadpole diagram 
which seems not to contribute to mass correction are taken into account. 
For simplicity, 
we have calculated 1-loop correction to Higgs mass via the bulk scalar field. 
Quantum corrections from the fields with other spin, 
such as the graviton, the vector field and the fermion, 
are easily obtained by 
multiplying the number of the degrees of freedom of physical polarization 
to the result from the bulk scalar field.

We have also calculated the effective potential for the Higgs field and 
obtained the finite mass in a systematic way. 
We have checked that the both results completely agree.

In Gauge-Higgs unification scenario, 
we can understand that the finite mass for Higgs field is generated 
by nontrivial appearance of Wilson loop. 
In Gravity-Gauge-Higgs unification scenario, however, 
it is not clear that a similar understanding can be applied 
since the naive correspondence to Wilson loop is not the line integral of 
the five-five component of the metric but the line integral of 
the Christoffel symbols, whose physical meaning has not been clarified so far.

Realistic model construction in this scenario is left for future investigations. 

\vspace{0.5cm}

\subsection*{Acknowledgment}
We would like to thank T. Inami for useful informative discussion. 
The authors would like to acknowledge the grant meeting 
``Gauge-Higgs Unification and Extra Dimension" held at Kobe University, 
Japan (January, 2004), where this work started.  
The work of C.S.L. was supported in part 
by the Grant-in-Aid for Scientific Research of the Ministry of Education, 
Science and Culture, No.15340078.  
N.M. is supported by Special Postdoctoral Researchers Program at RIKEN.

\end{document}